\documentclass[preprint,pra,showpacs,showkeys,tightenlines]{revtex4}
\usepackage{dcolumn}
\usepackage{bm}

\begin{document}

\title{Isotope shift calculations for atoms with one valence electron}

\author{J.C. Berengut}
\email{jcb@phys.unsw.edu.au}
\affiliation{School of Physics, University of New South Wales, Sydney 2052, Australia}
\author{V.A. Dzuba}
\affiliation{School of Physics, University of New South Wales, Sydney 2052, Australia}
\author{V.V. Flambaum}
\affiliation{School of Physics, University of New South Wales, Sydney 2052, Australia}

\date{16 May 2003}

\begin{abstract}

This work presents a method for the {\it ab~initio} calculation of isotope shift in atoms and ions with one valence electron above closed shells. As a zero approximation we use relativistic Hartree-Fock and then calculate correlation corrections. The main motivation for developing the method comes from the need to analyse whether different isotope abundances in early universe can contribute to the observed anomalies in quasar absorption spectra. The current best explanation for these anomalies is the assumption that the fine structure constant $\alpha$ was smaller at early epoch. We test the isotope shift method by comparing the calculated and experimental isotope shift for the alkali and alkali-like atoms Na, MgII, K, CaII and BaII. The agreement is found to be good. We then calculate the isotope shift for some astronomically relevant transitions in SiII and SiIV, MgII, ZnII and GeII.

\end{abstract}

\pacs{31.30.Gs, 31.15.Md, 31.25.Jf}
\keywords{isotope shift; mass shift; field shift; alkali metals}

\maketitle

\section{\label{sec:intro} Introduction}

Recent studies of quasar absorption spectra reveal a possible change in $\alpha$ since the early universe \cite{murphy1}. One of the possible major sources of systematic effects in these studies is that the isotopic abundance ratios in gas clouds in the early universe could be different to those on Earth. This may provide an alternative explanation for the observed variation in spectra \cite{murphy2}. In order to test this possibility it is necessary to have accurate values for the isotope shift (IS) for the relevant atomic transitions. Experimental data is available for only a very few of them. Therefore, accurate calculations are needed to make the most comprehensive analysis possible.

Another motivation for accurate isotope shift calculations comes from the possibility to use isotope shift to study atomic nuclei.  Comparing calculated and measured IS allows one to find the change in nuclear charge distribution from one isotope to another. Studying IS for these purposes has long history, for a good review and tables see e.g. \cite{heilig} or for more recent values in many atoms and ions \cite{aufmuth}. However any progress in the accuracy of calculations is of significant importance.

Isotope shift is also important in searching for small, charged black holes. The mass deficit of the observable universe (dark matter) may be explained by supposing the existence of non-vanishing ``elementary'' black holes of the Planck mass. Such black holes may have an electric charge, giving rise to the possibility of an atom made from electrons orbiting a positively charged black hole instead of a nucleus. Such black hole atoms would have spectra shifted with respect to usual nuclear atoms due to the effectively infinite mass and zero volume of the nucleus \cite{berengut}. This shift is simply an extreme example of the regular isotope shift, and can be calculated using the method described in this paper. A search for these spectra would verify the existence of elementary black holes, or any other stable, very heavy particles, e.g. ``strange~matter''.

In this paper, we develop a method for calculating the isotopic shift of atoms and ions that can be treated as a single electron above a closed shell core. These include the alkali metals, as well as other atoms and ions such as ZnII that are fairly well approximated by a single electron above a closed subshell core.

We test our technique by calculating IS for light alkaline atoms as well as for a relatively heavy ion, BaII. Since different contributions dominate in light and heavy atoms and good agreement with experiment has been obtained in both cases, one can confidently say that the technique has been reliably tested. We then apply the technique to calculate isotope shift for astronomically relevant transitions in MgII, ZnII, GeII, SiII and SiIV.

\section{\label{sec:method} Method}

The isotope shifts of atomic transition frequencies come from two sources: the finite size of the nuclear charge distribution (the ``volume'' or ``field'' shift), and the finite mass of the nucleus (see, e.g. \cite{sobelman}). The energy shift due to recoil of the nucleus is $(1/2M){\text p}_N^2~=~(1/2M)(\Sigma {\bf p}_i)^2$. Furthermore this ``mass shift'' is traditionally divided into the normal mass shift and the specific mass shift (SMS). The normal mass shift is given by the operator $(1/2M)\Sigma {\text p}_i^2$, which is easily calculated from the transition frequency. The SMS operator is $(1/M)\Sigma_{i<j}{\bf p}_i \cdot {\bf p}_j$ which is difficult to evaluate accurately.

The shift in energy of any transition in an isotope with mass number $A'$ with respect to an isotope with mass number $A$ can be expressed as
\begin{equation}
\label{eq:is}
\delta \nu^{A', A}
    = \left( k_{NMS} + k_{SMS} \right) \left( \frac{1}{A'} - \frac{1}{A} \right)
      + F \delta \langle r^2 \rangle ^{A', A} \ ,
\end{equation}
where the normal mass shift constant is
\begin{equation}
k_{NMS} = -\frac{\nu}{1822.888}
\end{equation}
and $\langle r^2 \rangle$ is the mean square nuclear radius. In this paper we develop a method for calculating the specific mass shift and field shift constants, $k_{SMS}$ and $F$ respectively. It is worth noting that in this paper we use the convention $\delta \nu^{A', A}~=~\nu^{A'} - \nu^{A}$.

To determine spectral shifts in black hole atoms relative to normal atoms, one must take $A' \rightarrow \infty$, and because the black hole has effectively zero volume $\delta \langle r^2 \rangle = -\langle r^2 \rangle = -\frac{3}{5} R_{\text{nuc}}^2$.

\subsection{\label{sec:sms} Specific mass shift}

It is well known that calculations using many-body perturbation theory (MBPT) in the residual Coulomb interaction give very poor convergence for atoms with many electrons. Therefore all order techniques were developed in earlier works to calculate energy levels, transition amplitudes, etc. (see, e.g. \cite{dzuba87,dzuba89a,dzuba89b,johnson91}). It is natural to expect that an all order technique is needed for the isotope shift as well. In the recent work by Safronova and Johnson (Ref.~\cite{johnson}) the SMS was calculated to third order of MBPT. Their results verify that there is very poor convergence for this operator. A first or second order calculation cannot even guarantee that the sign of the final result will be correct. A third order calculation gives the correct sign, however it is unknown how much fourth and higher order diagrams contribute.

Our method is to include the specific mass shift directly into an energy calculation from the very beginning. The SMS is a two-body operator ${\mathbf p}_1 \cdot {\mathbf p}_2$ and can be added to the Coulomb potential,
$\tilde Q = \frac{1}{|{\mathbf r}_1 - {\mathbf r}_2|} + \lambda {\mathbf p}_1 \cdot {\mathbf p}_2$
(see Appendix \ref{app:matrix}). The operator $\tilde Q$ replaces the Coulomb operator everywhere that it appears in an ``all order'' energy calculation that includes certain chains of diagrams to all orders. We vary the isotope shift scaling factor $\lambda$ and calculate the level energies $E$. The gradient of an $E$ vs. $\lambda$ graph then gives us the SMS matrix element, usually denoted $k_{SMS}$.

As a zero approximation we use relativistic Hartree-Fock (Hartree-Fock-Dirac) method. At the Hartree-Fock stage we include the isotope shift with the exchange potential and iterate to obtain self-consistent ``dressed'' wavefunctions. The SMS matrix element at this stage is roughly equivalent to the first-order and second-order contributions of the one-particle operator, plus higher-order corrections from the random-phase approximation in Ref.~\cite{johnson} ($P^{(1)} + S^{(2)}$ in their notation). It is interesting to note that $k_{SMS}$ at this stage does not give meaningful results, and can even be of the wrong sign (it is labelled as ``HF'' in Table~\ref{tab:k_sms_comp}). We need to include correlation corrections in order to obtain any reasonable accuracy.

We include correlation effects to second order of MBPT; this approach gives good accuracy for energies. The perturbation is the difference between the exact and Hartree-Fock Hamiltonians $V = H - H_{\text{HF}}$. The many-body corrections start in second order; all first order diagrams have been included in the self-consistent Hartree-Fock procedure. There are just four second order diagrams, illustrated for example in Ref.~\cite{dzuba87}. However now we replace the Coulomb operator in these diagrams with our two body operator $\tilde Q$, and use dressed Hartree-Fock wavefunctions as described above. This gives our final value of $k_{SMS}$ as listed in Table~\ref{tab:k_sms_comp}.

\begin{table}
\caption{\label{tab:k_sms_comp} Comparison of specific mass shift constants, $k_{SMS}$, obtained by including various contributions in the energy calculation (all in GHz.amu). Our values ``HF'' are approximately equivalent to the values labelled ``$P^{(1)} + S^{(2)}$'' here and in Ref.~\cite{johnson}. }
\begin{tabular}{llrrrrr}
\hline
\hline
Ion & State & \multicolumn{2}{c}{Ref.~\cite{johnson}} & & \multicolumn{2}{c}{This work} \\
\cline{3-4} \cline{6-7}
    &       & $P^{(1)} + S^{(2)}$ & Final & & HF & Final \\
\hline
NaI  & $3{\text s}$       & -54 &  54 & &  -52 &   69 \\
     & $3{\text p}_{1/2}$ & -67 & -43 & &  -68 &  -40 \\
     & $3{\text p}_{3/2}$ & -67 & -43 & &  -67 &  -39 \\
\\
MgII & $3{\text s}$       &     &   38 & & -171 &   83 \\
     & $3{\text p}_{1/2}$ &     & -324 & & -408 & -296 \\
     & $3{\text p}_{3/2}$ &     & -323 & & -402 & -290 \\
\hline
\hline
\end{tabular}
\end{table}

An estimate of the size of neglected higher order diagrams can be obtained by calculating a new value of $k_{SMS}$ that includes some higher order chains of diagrams. We create an operator $\Sigma$ for the second order correlation effects in each partial wave, defined by
\begin{equation}
\label{eq:sigma_def}
\delta E_{n}^{(2)} = \langle n | \Sigma | n \rangle \ .
\end{equation}
This ``sigma'' operator is then added to the exchange potential in the Hartree-Fock Hamiltonian for the valence electron, $\tilde H = H_{\text{HF}} + \Sigma$. Thus we calculate the single electron Brueckner orbital. We also include a scaling factor, $f$, with this sigma operator ($\Sigma \rightarrow f \Sigma$) in order to fit the experimental energy. The rescaling of $\Sigma$ simulates some higher order correlation corrections that were omitted in our calculation. Including $\Sigma$ in our Hartree-Fock calculation allows us to obtain another value for $k_{SMS}$, and the difference between this new value and the old second-order value gives us an estimate of the error.

In Table~\ref{tab:k_sms_comp} we compare our results to those obtained by Safronova and Johnson (Ref.~\cite{johnson}) who calculated the isotope shift to third order in standard MBPT. Our values are significantly different, in particular the MgII 3s level shift is more than twice that given in Ref.~\cite{johnson} and also \cite{veseth}, which are broadly in agreement with each other. However we find that this difference is not so large in the $3\text{p} - 3\text{s}$ transition due to cancellations of higher order terms between the two levels. We compare our final results and those of Ref.~\cite{johnson} with experiment in Table~\ref{tab:total_sms_comp}. Agreement is at the level of ~1\%, which is much smaller than our error estimates.

Our method includes many-body diagrams that are different to Ref.~\cite{johnson}, including some chains of diagrams in all orders. It is also arguably simpler to implement since it doesn't require term-by-term calculation of a large number of matrix elements.

\begin{table}
\caption{\label{tab:total_sms_comp} Comparison of the specific mass shift of transitions in Na and MgII with experiment.}
\begin{tabular}{llrrr}
\hline
\hline
Isotopes & Transition & \multicolumn{3}{c}{SMS (MHz)} \\
\cline{3-5}
         &            & Ref.~\cite{johnson} & This work & Exp.\footnotemark[1] \\
\hline
$^{23 - 22}$Na 
   & $3{\text p}_{1/2} - 3{\text s}$ & 192 & 214(48) & 215(1) \\
   &                                 &     &         & 214(2) \\
   & $3{\text p}_{3/2} - 3{\text s}$ & 192 & 212(48) & 214    \\
$^{26 - 24}$MgII 
   & $3{\text p}_{3/2} - 3{\text s}$ & 1157 & 1196(18) & 1207(100) \\
\hline
\hline
\end{tabular}
\footnotetext[1]{these values are extracted from IS experiments in Table \ref{tab:sms_experiments}}
\end{table}

\subsection{\label{sec:field} Field shift}

Our method for calculating the field shift (also called the volume shift) is similar to that used for the SMS calculation. We add a perturbation due to the field shift to the nuclear potential, and then calculate the energy directly. The field shift potential is
\begin{equation}
\delta U(r) = \lambda \left(\, U(R + \delta R, r) - U(R, r)\, \right)
\end{equation}
where $R$ is the nuclear radius and $U(R, r)$ is the nuclear potential. To obtain the change in energy of a state due to the field shift, we add this potential to the nuclear potential in our Hartree-Fock calculations. By introducing a scaling factor $\lambda$ we can check linearity and increase the size of the effect. To calculate the field shift constant, we take the gradient of a level energy $E$ vs. $\lambda$ graph and extract $F$ as
\begin{equation}
F = \frac{\delta E_V}{\delta \langle r^2 \rangle}
     = \frac{5}{3} \frac{1}{2 R \, \delta R} \frac{dE}{d\lambda} \ .
\end{equation}
Note that the field shift constant, $F$, is defined here as having opposite sign to the definition in some previous papers, e.g. Ref.~\cite{johnson} and~\cite{MP92}. This equation assumes that the nuclear charge distribution can be approximated as a uniformly charged sphere of radius $R$.

Higher order correlation effects are smaller in the field shift than in the specific mass shift, and are localised at the nucleus. We can include them easily by creating Brueckner orbitals (solutions of the modified Hamiltonian, $H = H_{\text{HF}} + \Sigma$) using a second order sigma operator, defined in Equation~\ref{eq:sigma_def}. The field shift is proportional in first order to the square of the wavefunction at the nucleus. Hence we can include higher order effects quite simply by multiplying the matrix element by the square of the ratio of the Brueckner wavefunction to the Hartree-Fock wavefunction at the nucleus,
\begin{equation}
\frac{F_{{\text{HF}} + \Sigma}}{F_{\text{HF}}} 
= \left| \frac{\psi_{{\text{HF}} + \Sigma}(0)}{\psi_{\text{HF}}(0)} \right|^2 \ .
\end{equation}

We tested this in BaII, as well as in lighter atoms, because in barium the field shift dominates over the mass shift, and there is a lot of experimental data to compare with. Our results for field shift constants in BaII were found to be consistent within a few percent of the previous theoretical work by M\aa rtensson-Pendrill (Table~\ref{tab:BaII_level}). In the same paper, Ref.~\cite{MP92}, they say that they have underestimated the $6{\text p}_{1/2}$ field shift constant by around 7\%, based on the difference between experimental and theoretical calculations of the hyperfine $A$ constant. Also, their $6{\text s}$ constant is said to be overestimated, leading to a corrected value of $F_{6{\text p}_{3/2}-6{\text s}}=-4.20(13)$~GHz/fm$^2$. These corrected values are in better agreement with the {\it ab initio} values obtained in this work ($F_{6{\text p}_{3/2}-6{\text s}}=-4.076$~GHz/fm$^2$).

\begin{table}
\caption{\label{tab:BaII_level} Level field shifts in BaII states. Due to a difference in the definition of $F$, the values calculated in Ref.~\cite{MP92} have been presented here with opposite sign.}
\begin{tabular}{ldrr}
\hline
\hline
State    & \multicolumn{1}{c}{Energy}      & \multicolumn{2}{c}{$F$ (MHz/fm$^2$)} \\
\cline{3-4}
         & \multicolumn{1}{c}{(cm$^{-1}$)} & Ref.~\cite{MP92} & This work \\
\hline
$6{\text s}$       & -80686.87 & $4096$   & $3851$   \\
$6{\text p}_{1/2}$ & -60425.31 & $-111.1$ & $-150.1$ \\
$6{\text p}_{3/2}$ & -58734.45 & $-242.6$ & $-225.4$ \\
$5{\text d}_{3/2}$ & -75813.02 &          & $-1223$  \\
$5{\text d}_{5/2}$ & -75012.05 &          & $-1148$  \\
\hline
\hline
\end{tabular}
\end{table}

\begin{table}
\caption{\label{tab:BaII_transitions} Calculated SMS and field shift constants in BaII transitions.}
\begin{tabular}{lrrr}
\hline
\hline
Transition & Wavelength (nm) & $k_{SMS}$ (GHz.amu) & $F$ (MHz/fm$^2$)  \\
\hline
$6{\text p}_{1/2} - 6{\text s}$       &  493 &  105 & -4001 \\
$6{\text p}_{3/2} - 6{\text s}$       &  455 &  257 & -4077 \\
$5{\text d}_{5/2} - 6{\text s}$       & 1762 & -550 & -4999 \\
$6{\text p}_{1/2} - 5{\text d}_{3/2}$ &  650 &  653 &  1073 \\
$6{\text p}_{3/2} - 5{\text d}_{3/2}$ &  585 &  805 &   997 \\
$6{\text p}_{3/2} - 5{\text d}_{5/2}$ &  614 &  807 &   922 \\
\hline
\hline
\end{tabular}
\end{table}

Using King plots \cite{king} we can extract the ratios of field shift constants for different transitions, provided we have experimental data for a number of different isotopes. In Table~\ref{tab:BaII_ratios} we compare our calculated values of these ratios with those obtained by combining the data in several different experiments and transitions. Our values were found to be consistent with this experiment to within 5\%. We have used a simple weighted least squares fit to obtain an experimental value for the ratio $F_{455}/F_{614}$. A two point formula was used for ratios involving the 1762~nm transition as experimental data exists only for the $\delta \nu^{134, 138}$ and $\delta \nu^{136, 138}$ isotope shifts \cite{zhao}. Other ratios were extracted by the groups that performed the experiments, with much higher accuracy.

\begin{table}
\caption{\label{tab:BaII_ratios} Ratios of field shift constants in BaII states. In the second column we list the measured values, obtained using King plots. In some cases we obtain the ratio ourselves by combining the results of two separate studies. }
\begin{tabular}{rldd}
\hline
\hline
\multicolumn{2}{c}{Transitions} & \multicolumn{2}{c}{$F_{\alpha}/F_{\beta}$} \\
 (\, $\alpha$ & $/\ \beta$\, )  & \multicolumn{1}{c}{This work} & \multicolumn{1}{c}{Experiment} \\
\hline
$6{\text p}_{1/2} - 6{\text s}$ & $/\ 6{\text p}_{3/2} - 6{\text s}$
        &  0.982 & 0.975(3)\footnotemark[1] \\
$6{\text p}_{3/2} - 6{\text s}$ & $/\ 6{\text p}_{3/2} - 5{\text d}_{5/2}$
        & -4.42  & -4.50(6)\footnotemark[2]\footnotemark[3] \\
$6{\text p}_{3/2} - 6{\text s}$ & $/\ 5{\text d}_{5/2} - 6{\text s}$
        &  0.816 &  0.82(4)\footnotemark[2]\footnotemark[4] \\
$5{\text d}_{5/2} - 6{\text s}$ & $/\ 6{\text p}_{3/2} - 5{\text d}_{5/2}$
        & -5.42  &  -5.5(3)\footnotemark[3]\footnotemark[4] \\
$6{\text p}_{1/2} - 5{\text d}_{3/2}$ & $/\ 6{\text p}_{3/2} - 5{\text d}_{3/2}$
        &  1.076 & 1.087(4)\footnotemark[3] \\
$6{\text p}_{3/2} - 5{\text d}_{5/2}$ & $/\ 6{\text p}_{3/2} - 5{\text d}_{3/2}$
        &  0.925 & 0.961(3)\footnotemark[3] \\
$6{\text p}_{3/2} - 5{\text d}_{3/2}$ & $/\ 6{\text p}_{3/2} - 6{\text s}$
        & -0.245 & -0.2312(6)\footnotemark[3] \\
\hline
\hline
\end{tabular}
\footnotetext[1]{Wendt {\it et~al.} \cite{wendt84}}
\footnotetext[2]{Wendt {\it et~al.} \cite{wendt88}}
\footnotetext[3]{Villemoes {\it et~al.} \cite{villemoes}}
\footnotetext[4]{Zhao {\it et~al.} \cite{zhao}}
\end{table}

\subsection{\label{sec:alkaline} Alkaline ions}

We compare our results with experimental data for alkaline ions in Table~\ref{tab:sms_experiments}. In alkaline ions it is more valuable to compare only the specific mass shift with those extracted from experiment, than to compare the entire isotope shift. This is because the mass shift dominates strongly in these ions, and also because the SMS is generally considered more difficult to calculate. We have removed the field shift and the normal mass shift from the experimental values of the isotopic shift in order to obtain an experimental value for the specific mass shift. The field shift values used in Table~\ref{tab:sms_experiments} were calculated using the above method. While our calculation of $F$ has been shown to be good, the field shift also depends on having knowledge of $\delta \langle r^2 \rangle$ for the relevant isotopes. 

\begingroup
\squeezetable
\begin{table*}
\caption{\label{tab:sms_experiments} Comparison of experimental values of the specific mass shift with calculated theoretical values. The experimental values were extracted by subtracting the NMS and field shift (FS) from the experimental IS. }
\begin{tabular}{llrrrrrr}
\hline
\hline
Isotopes & Transition & Energy    & IS (exp) & NMS   & FS
                                                 & \multicolumn{2}{c}{SMS (MHz)} \\
\cline{7-8}
         &            & cm$^{-1}$ & (MHz)    & (MHz) & (MHz) & This work & Exp. \\
\hline
$^{23 - 22}$Na
   & $3{\text p}_{1/2} - 3{\text s}$ & 16956.18 & 758.5(7)\footnotemark[1] 
                                                          & 551 & -8    & 214(48) & 215(1) \\
   &                                 &          & 756.9(1.9)\footnotemark[2]
                                                          &     &       &         & 214(2) \\
   & $3{\text p}_{3/2} - 3{\text s}$ & 16973.38 & 757.72(24)\footnotemark[3]
                                                          & 552 & -8    & 212(48) & 214    \\
$^{26 - 24}$MgII
   & $3{\text p}_{3/2} - 3{\text s}$ & 35760.97 & 3050(100)\footnotemark[4]
                                                          & 1185 & -42 & 1196(18)& 1207(100)\\
$^{41 - 39}$K
   & $4{\text p}_{1/2} - 4{\text s}$ & 12985.17 & 235.25(75)\footnotemark[5]
                                                          & 267 & -13(5) & -32(21) & -19(6) \\
   & $4{\text d}_{1/2} - 4{\text s}$ & 27398.11 & 585(9)\footnotemark[6]
                                                          & 564 & -13(5) &  20(30) & 34(13) \\
$^{43 - 40}$CaII
   & $4{\text p}_{1/2} - 4{\text s}$ & 25191.54 & 706(42)\footnotemark[7]
                                                          & 723 & -36(3) &  22(1) & 19(45)  \\
   &                                 &          & 672(9)\footnotemark[8]
                                                          &     &        &        & -15(11) \\
   &                                 &          & 685(36)\footnotemark[9]
                                                          &     &        &        & -2(39)  \\
   & $4{\text p}_{3/2} - 4{\text s}$ & 25414.43 & 713(31)\footnotemark[7]
                                                          & 729 & -36(3) &  -5(1) & 20(34)  \\
   &                                 &          & 677(19)\footnotemark[8]
                                                          &     &        &        & -16(22) \\
   &                                 &          & 685(36)\footnotemark[9]
                                                          &     &        &        & -8(39)  \\
   & $3{\text d}_{3/2} - 4{\text s}$ & 13650.21 & 4180(48)\footnotemark[7]
                                                          & 392 & -47(4) & 3502(217) & 3835(52) \\
   & $3{\text d}_{5/2} - 4{\text s}$ & 13710.90 & 4129(10)\footnotemark[7]
                                                          & 393 & -47(4) & 3487(215) & 3783(14) \\
   & $4{\text p}_{1/2} - 3{\text d}_{3/2}$ & 11541.33 & -3464.3(3.0)\footnotemark[10]
                                                          & 331 & 12(1) & -3479(218) & -3807(4) \\
   &                                 &          & -3483(40)\footnotemark[7]
                                                          &     &       &            & -3826(41)\\
   & $4{\text p}_{3/2} - 3{\text d}_{3/2}$ & 11764.22 & -3462.4(2.6)\footnotemark[10]
                                                          & 337 & 12(1) & -3507(217) & -3811(4) \\
   &                                 &          & -3446(20)\footnotemark[7]
                                                          &     &       &            & -3795(21)\\
   & $4{\text p}_{3/2} - 3{\text d}_{3/2}$ & 11703.53 & -3465.4(3.7)\footnotemark[10]
                                                          & 336 & 12(1) & -3492(216) & -3813(5) \\
   &                                 &          & -3427(33)\footnotemark[7]
                                                          &     &       &            & -3774(34)\\
\hline
\hline
\end{tabular}
\footnotetext[1]{Pescht {\it et~al.} \cite{pescht}}
\footnotetext[2]{Huber {\it et~al.} \cite{huber}}
\footnotetext[3]{Gangrsky {\it et~al.} \cite{gangrsky}}
\footnotetext[4]{Drullinger {\it et~al.} \cite{drullinger}}
\footnotetext[5]{Touchard {\it et~al.} \cite{touchard}}
\footnotetext[6]{H\"{o}rb\"{a}ck {\it et~al.} \cite{horback}}
\footnotetext[7]{Kurth {\it et~al.} \cite{kurth}}
\footnotetext[8]{extracted from M\aa rtensson-Pendrill {\it et~al.} \cite{MP92_CaII}}
\footnotetext[9]{Maleki and Goble\cite{maleki}}
\footnotetext[10]{N\"{o}rtersh\"{a}user {\it et~al.} \cite{nortershauser}}
\end{table*}
\endgroup

For Na we use the value quoted in Ref.~\cite{johnson} of $\delta \langle r^2 \rangle^{23, 22}~=~0.205(3)~{\text{fm}}^2$. This value is only from an empirical fit, and shouldn't be trusted too far. The field shift is very small in this atom, so the errors don't matter too much. For MgII we have the used value $\delta \langle r^2 \rangle^{26, 24} = 0.55~{\text{fm}}^2$ from another empirical fit, the equation $R_{\text{nuc}} = 1.1 A^{1/3}~{\text{fm}}$. This is very poor, but in this case the field shift is small even in relation to the error in the experimental isotope shift. In Table~\ref{tab:sms_experiments} we have not included an error contribution for the field shift in either of these atoms, since we really don't know how accurate these approximations are.

The values of $\delta \langle r^2 \rangle$ are known for K and CaII from muonic x-ray experiments, allowing us to calculate the field shift much more accurately. This is fortunate because the SMS is relatively small for the $\text{p}-\text{s}$ transitions in these atoms, and hence the field shift plays a much larger role. We use the values $\delta \langle r^2 \rangle^{41, 39} = 0.117(40)~{\text{fm}}^2$ for K from Ref.~\cite{wohlfahrt}, and $\delta \langle r^2 \rangle^{43, 40} = 0.1254(32)~{\text{fm}}^2$ for CaII from Ref.~\cite{palmer}. In CaII the change in mean square nuclear radius is given to high precision, so we have included an additional error of 5\% in the field shift that comes from the constant $F$. This is a pessimistic estimate of error based on the accuracy we achieved calculating $F$ for transitions in BaII.

Table~\ref{tab:sms_experiments} shows that our method can reliably calculate the isotope shift in alkaline atoms, including those transitions with a large specific mass shift.

\section{\label{sec:results} Results}

We have shown that our method works in atoms for which we have available experimental data (Section~\ref{sec:method}). In Table~\ref{tab:constants} we tabulate values for the mass and field shift constants for some astronomically useful transitions. We have not given errors for $F$, however we can say that they are less than 5\% based on comparison of calculation with experiment in BaII.

\begin{table}
\caption{\label{tab:constants} Mass and field shift constants for some useful transitions.}
\begin{tabular}{llrrr}
\hline
\hline
Ion   & Transition & $F$          & $k_{NMS}$ & $k_{SMS}$ \\
      &            & (MHz/fm$^2$) & (GHz.amu) & (GHz.amu) \\
\hline
MgII  & $3{\text p}_{1/2} - 3{\text s}$ & -127 &  -587 &  -373(12) \\
      & $3{\text p}_{3/2} - 3{\text s}$ & -127 &  -588 &  -373(6)  \\
SiII  & $4{\text s} - 3{\text p}_{1/2}$ &  171 & -1077 &  1257(29) \\
      & $4{\text s} - 3{\text p}_{3/2}$ &  171 & -1072 &  1243(28) \\
SiIV  & $3{\text p}_{1/2} - 3{\text s}$ & -484 & -1172 & -1535(11) \\
      & $3{\text p}_{3/2} - 3{\text s}$ & -485 & -1180 & -1505(7)  \\
ZnII  & $4{\text p}_{1/2} - 4{\text s}$ &-1596 &  -797 & -1310(69) \\
      & $4{\text p}_{3/2} - 4{\text s}$ &-1596 &  -812 & -1266(69) \\
GeII  & $5{\text s} - 4{\text p}_{1/2}$ & 1088 & -1026 &  1046(69) \\
      & $5{\text s} - 4{\text p}_{3/2}$ & 1083 &  -997 &   960(62) \\
\hline
\hline
\end{tabular}
\end{table}

In Table~\ref{tab:total_values} we present the results of isotope shift calculations between common isotopes of astronomically important ions. We have used the IS constants presented in Table~\ref{tab:constants} with Equation~\ref{eq:is} in order to calculate the isotope shift between particular isotopes. Just before submission of this paper, results of measurements for ZnII were brought to our attention \cite{matsubara}. These results matched our prediction extremely well.

In GeII and SiII, the specific mass shift cancels the normal mass shift entirely, making the field shift, and hence $\delta \langle r^2 \rangle$, important. We have just used the empirical formula $R_{\text{nuc}} = 1.1 A^{1/3}~{\text{fm}}^2$ to obtain values of $\delta \langle r^2 \rangle$ in these ions. This is extremely rough, and although it seems to work in ZnII to within a 20\% accuracy based on the experimental data given, we really don't know if this holds for Si and Ge at all. In the SiIV transitions presented, it is less important to have good values for the field shift because there is no cancellation between the NMS and SMS.

\begin{table}
\caption{\label{tab:total_values} Summary of isotope shift values for astronomically relevant alkali-like ions. The experimental value for the $4{\text p}_{3/2} - 4{\text s}$ transition in ZnII is 676(6)~MHz as quoted in \cite{matsubara}. We have presented two errors, the first is our uncertainty in $k_{SMS}$ and the second is the uncertainty in our field shift, which is mainly due to lack of knowledge of $\delta \langle r^2 \rangle$. A negative shift means that the sign is opposite to the normal mass shift. }
\begin{tabular}{llrr}
\hline
\hline
Isotopes & Transition & \multicolumn{1}{c}{Energy} & \multicolumn{1}{c}{Isotope Shift} \\
         &            & \multicolumn{1}{c}{(cm$^{-1}$)} & \multicolumn{1}{c}{(MHz)}    \\
\hline
$^{30 - 28}\text{SiII}$ & $4{\text s} - 3{\text p}_{1/2}$ & 65495.1 & -375(70)(11) \\
                        & $4{\text s} - 3{\text p}_{3/2}$ & 65208.1 & -351(67)(11) \\
$^{30 - 28}\text{SiIV}$ & $3{\text p}_{1/2} - 3{\text s}$ & 71289.6 & 6294(26)(31) \\
                        & $3{\text p}_{3/2} - 3{\text s}$ & 71749.9 & 6241(18)(31) \\
$^{66 - 64}\text{ZnII}$ & $4{\text p}_{1/2} - 4{\text s}$ & 48480.6 &  653(32)(78) \\
                        & $4{\text p}_{3/2} - 4{\text s}$ & 49354.4 &  632(33)(79) \\
$^{74 - 70}\text{GeII}$ & $5{\text s} - 4{\text p}_{1/2}$ & 62402.4 &  491(53)(101) \\
                        & $5{\text s} - 4{\text p}_{3/2}$ & 60635.3 &  533(48)(101) \\
\hline
\hline
\end{tabular}
\end{table}

\section{\label{conclusion} Conclusion}

We have presented a method for the calculation of the isotope shift in atoms and ions that can be approximated as having one valence electron above a closed shell. Our results are shown to be in good agreement with isotope shift experiments in both light and heavy atoms, which are good tests for the mass shift and field shift respectively.

We have used the method to predict values of the isotope shift in astronomically relevant transitions in SiII and SiIV, MgII, ZnII and GeII. Recent experiments measured the isotope shift of a ${\text p} - {\text s}$ transition in ZnII, and the results were in excellent agreement with our prediction. These values are needed in order to examine systematic effects in observations of spectral line shifts in quasar absorption spectra that suggest a variation in $\alpha$ \cite{murphy2}. Our calculations could help provide another explanation for the observed shifts, depending on whether isotopic abundances were different in gas clouds in the early universe. Alternatively, our calculations may strengthen the arguments in support of a varying $\alpha$.

Further work needs to be done in order to obtain the isotope shift for more complex atoms with more than one electron in their outer shell. The general method of including the isotope shift operator with the two-body Coulomb operator may be used in such cases, although the energy calculation itself is more complicated. 

\begin{acknowledgments}
The authors would like to thank Jacinda Ginges for useful discussions. We would also like to thank Bruce Warrington for bringing Reference~\cite{matsubara} to our attention. This work was supported by the Australian Research Council.
\end{acknowledgments}

\newpage
\appendix*
\section{\label{app:matrix} Matrix element of the two-body operator}

The two-body operator used is this work is the sum of the Coulomb interaction
operator and the ``rescaled'' SMS operator (atomic units):
\begin{equation}
	\tilde Q = \frac{1}{|{\mathbf r}_1 - {\mathbf r}_2|} + \lambda
	 {\mathbf p}_1 \cdot {\mathbf p}_2 \equiv \sum_k \tilde Q_k,
\label{a1}
\end{equation}
where $\lambda$ is the scaling factor, ${\mathbf p} = -i\nabla$ is electron 
momentum, and
\begin{equation}
	\tilde Q_k = \frac{4\pi}{2k+1}\frac{r_<^k}{r_>^{k+1}}
	Y_{k}({\mathbf n}_1)Y_{k}({\mathbf n}_2) + \lambda\, 
	{\mathbf p}_1 \cdot {\mathbf p}_2\, \delta_{k1}.
\label{a2}
\end{equation}
We use the following form for the single-electron wave function
\begin{eqnarray}
	\psi({\mathbf r})_{jlm} = \frac{1}{r}
 	\left( \begin{array}{c}
	f(r) \Omega({\mathbf n})_{jlm} \\ i 
	\alpha g(r) \tilde{\Omega}({\mathbf n})_{jlm}
	\end{array}
	\right).
\label{psi}
\end{eqnarray}
Here $\alpha = 1/137.036$ is the fine structure constant, and 
$\tilde{\Omega}({\mathbf n})_{jlm} = -(\vec \sigma \cdot {\mathbf n}) 
{\Omega}({\mathbf n})_{jlm}$.

The matrix element of operator (\ref{a2}) with wave functions (\ref{psi})
has the form
\begin{equation}
	\langle \psi_1({\mathbf r}_1)\psi_2({\mathbf r}_2)|\tilde Q_k|
	\psi_3({\mathbf r}_1)\psi_4({\mathbf r}_2) \rangle =
	C_k (R_k - \lambda P_1 \delta_{k1}),
\label{a3}
\end{equation}
where the angular factor $C_k$ is the same for both operators
\begin{eqnarray}
	& C_k = (-1)^{q+m_1+m_2} \left( 
	\begin{array}{ccc} j_1 & k & j_3 \\ -m_1 & q & m_3 \end{array}
	\right) \left(
	\begin{array}{ccc} j_2 & k & j_4 \\ -m_2 & -q & m_4 \end{array} 
	\right) \nonumber \\ 
\label{a4}
	& \times (-1)^{j_1+j_2+j_3+j_4+1}
	\sqrt{(2j_1+1)(2j_2+1)(2j_3+1)(2j_4+1)} \\*
	& \times \left(
	\begin{array}{ccc} j_1 & j_3 & k \\ \frac{1}{2} & -\frac{1}{2} & 0 
	\end{array}  \right)  \left(
	\begin{array}{ccc} j_2 & j_4 & k \\ \frac{1}{2} & -\frac{1}{2} & 0 
	\end{array}  \right) \xi(l_1+l_3+k)\xi(l_2+l_4+k) \nonumber, \\*
	& \xi(x) = \left\{ \begin{array}{ll} 1, & \text{if} ~x~ 
	\text{is even}, \\*
	0, & \text{if} ~x~ \text{is odd} \end{array} \right. \nonumber,
\end{eqnarray}
$R_k$ is radial Coulomb integral
\begin{equation}
	R_k = \int_0^{\infty}\frac{r_<^k}{r_>^{k+1}}
	(f_1(r_1)f_3(r_1)+\alpha^2 g_1(r_1)g_3(r_1))
	(f_2(r_2)f_4(r_2)+\alpha^2 g_2(r_2)g_4(r_2)) dr_1dr_2,
\label{a5}
\end{equation}
while $P_1$ is radial matrix element of the SMS operator
\begin{eqnarray}
	& P_1 = p_{13}p_{24}, \nonumber \\
\label{a6}
	& p_{ab} = A_{ab} \delta_{l_a l_b+1} + B_{ab} \delta_{l_a l_b-1}, \\*
	& A_{ab} = \int_0^{\infty}f_a(\frac{d}{dr}-\frac{l_a}{r})f_b dr,
	\nonumber \\*
	& B_{ab} = \int_0^{\infty}f_a(\frac{d}{dr}+\frac{l_b}{r})f_b dr.
	\nonumber
\end{eqnarray}

\end{document}